\documentclass{aa}

\usepackage{txfonts}
\usepackage{amssymb}
\usepackage{amsmath}
\usepackage{graphicx}
\usepackage{natbib}
\usepackage{color}
\usepackage{soul}

\begin{document}

\title{Flux decay during thermonuclear X-ray bursts analysed with the dynamic power-law index method}  

\titlerunning{X-ray burst decays from dynamic power-law cooling index method}

\author{J.\,Kuuttila\inst{1}
\and J.~J.~E.~Kajava\inst{2,1,3}
\and J.\,N\"attil\"a\inst{1,4}
\and S.~E.~Motta\inst{5}
\and C. S\'anchez-Fern\'andez\inst{3}
\and E. Kuulkers\inst{3,6}
\and A. Cumming\inst{7}
\and J.~Poutanen\inst{1,4,8} 
}

\institute{Tuorla Observatory, Department of Physics and Astronomy, University of Turku, V\"ais\"al\"antie 20, FI-21500 Piikki\"o, Finland
\and Finnish Centre for Astronomy with ESO (FINCA), University of Turku, V\"{a}is\"{a}l\"{a}ntie 20, FIN-21500 Piikki\"{o}, Finland
\and European Space Astronomy Centre (ESA/ESAC), Science Operations Department, 28691 Villanueva de la Ca\~nada, Madrid, Spain \and
Nordita, KTH Royal Institute of Technology and Stockholm University, Roslagstullsbacken 23, SE-10691 Stockholm, Sweden \and 
University of Oxford, Department of Physics, Astrophysics, Denys Wilkinson Building, Keble Road, Oxford OX1 3RH, UK \and
ESA/ESTEC, Keplerlaan 1, 2201 AZ Noordwijk, The Netherlands \and
Department of Physics and McGill Space Institute, McGill University, 3600 rue University, Montreal, QC H3A2T8, Canada
\and 
Kavli Institute for Theoretical Physics, 
University of California, Santa Barbara, CA 93106, USA 
}


\abstract{The cooling of type-I X-ray bursts can be used to probe the nuclear burning conditions in neutron star envelopes. 
The flux decay of the bursts has been traditionally modelled with an exponential, even if theoretical considerations predict power-law-like decays. 
We have analysed a total of 540 type-I X-ray bursts from five low-mass X-ray binaries observed with the \textit{Rossi X-ray Timing Explorer}.
We grouped the bursts according to the source spectral state during which they were observed (hard or soft), flagging those bursts that showed signs of photospheric radius expansion (PRE). 
The decay phase of all the bursts were then fitted with a dynamic power-law index method. 
This method provides a new way of probing the chemical composition of the accreted material.
Our results show that in the hydrogen-rich sources the power-law decay index is variable during the burst tails and that simple cooling models qualitatively describe the cooling of presumably helium-rich sources 4U 1728--34 and 3A 1820--303. 
The cooling in the hydrogen-rich sources 4U 1608--52, 4U 1636--536, and GS 1826--24, instead, is clearly different and depends on the spectral states and whether PRE occurred or not. 
Especially the hard state bursts behave differently than the models predict, exhibiting a peculiar rise in the cooling index at low burst fluxes, which suggests that the cooling in the tail is much faster than expected.
Our results indicate that the drivers of the bursting behaviour are not only the accretion rate and chemical composition of the accreted material, but also the cooling that is somehow linked to the spectral states. 
The latter suggests that the properties of the burning layers deep in the neutron star envelope might be impacted differently depending on the spectral state.
}


\maketitle

\section{Introduction}\label{sec:intro}

X-ray bursts were first discovered in 1975 by \citet{Grindlay76}, who observed bright flashes coming from a low-mass X-ray binary (LMXB), 3A 1820--303 (but see also \citealt{Belian72,Babushkina75}). Since then, type-I thermonuclear X-ray bursts have been observed from more than 100 LMXBs \citep{Galloway08}\footnote{See also http://burst.sci.monash.edu/sources\\ and https://personal.sron.nl/~jeanz/bursterlist.html}. X-ray bursts are triggered by the accreted matter from the companion star that piles up on the neutron star (NS) surface until the pressure is sufficiently high for the ignition of the thermonuclear fusion \citep{Bildsten98}. 
The burning is usually unstable and most of the accumulated material burns in a matter of seconds (for a review, see \citealt{Lewin93,Strohmayer06}). 
Some of these bursts are so energetic that they reach the local Eddington limit, causing the atmosphere of the star to momentarily expand during the burning. 
These are often dubbed as the photospheric radius expansion (PRE) bursts.

The bursting behaviour may be affected significantly by the spectral (or accretion) state of the binary (see e.g. \citealt{Galloway08, Kajava14}, and references therein). 
Many bursting NSs in LXMBs alternate between so-called hard and soft spectral states \citep{MDFM14}, and X-ray bursts can occur while the LMXB is in either of those two states or during state transitions (i.e. the intermediate state). 
In the hard state the energy spectrum is dominated by Comptonised emission (see e.g. \citealt{DGK07}). 
The optically thick accretion disc is likely truncated at some distance from the NS, and a hot, optically thin, and geometrically thick inner flow channels the accreted gas onto the NS. 
In the soft state the accretion disc is thought to reach the surface of the NS and two thermal components dominate the energy spectrum. 
The accretion disc dominates the X-ray emission in soft X-rays, and a hotter optically thick spreading layer that is thought to form in the NS-accretion disc boundary \citep{IS99,SP06,IS10} emits harder X-rays.
Importantly, according to \citet{IS10} a significant fraction of the accreted energy can be dissipated in the burning layers, and it is thus possible that the accretion state is a key driver of the bursting properties.

Observationally, type-I X-ray bursts are characterised by a rapid rise of the X-ray flux by a factor of up to $\sim$ 100 with respect to the persistent emission, after which the X-ray flux slowly decreases back to the levels prior to the burst, usually within a minute or two. 
The flux decay has traditionally been modelled with exponential functions because early attempts to explain the bursts predicted an exponential decay, and these models matched well the decay of the first bursts observed \citep{Grindlay76b}. 
On the other hand, a simple physical consideration of one-dimensional radiative heat transfer implies a power-law-like dependency for the observed flux, and \citet{intZand14} showed that the single power law successfully explains the flux decays for a subset of 35 X-ray bursts from 14 sources. 

In both cases (exponential decays or power-law models) the decay rate was assumed to be constant throughout the cooling tail \citep{intZand14}. 
However, cooling models developed by \citet{Cumming04} predict that the power-law decay index should decrease while the flux drops. 
These models were initially constructed for superbursts, and later \citet{intZand14} adapted these models for shallower column depths with an energy release of 1.6 MeV per nucleon, as expected from complete helium burning to evaluate the behaviour of normal bursts.
These models assume an instantaneous energy injection and then follow the subsequent thermal evolution of the burning layer.
They predict a smooth decrease of the power-law index, from over 2 towards 1 (or even lower depending on the parameters) with decreasing flux (see figure 6 of \citealt{intZand14}).
Thus, it is clear that the full cooling behaviour cannot be fully described with only one or two power laws fitted to the burst tail nor is an exponential decay model sufficient to capture all of the relevant physics.

In this paper we study how the predicted decay rate changes during type-I X-ray bursts of five LMXBs using a dynamic power-law index method, where a power law is fitted to the flux-time evolution within a moving time window.
As a first step of this kind of dynamic decay modelling, we compare the observations to the aforementioned simple models developed in \citet{Cumming04} and in \citet{intZand14}.
The five systems have hydrogen-rich and hydrogen-poor accretors, which also allows us to study how the chemical composition affects details of the cooling.
In addition, we study the burst decay behaviour as a function of spectral state to see if the accretion geometry affects the conditions in the burning depths.
Finally, we investigate whether the Eddington limited PRE bursts have different cooling behaviour compared to the fainter non-PRE bursts.

\begin{figure*} [!ht]
\begin{center}
\includegraphics[width=0.99\textwidth, height=2.5in]{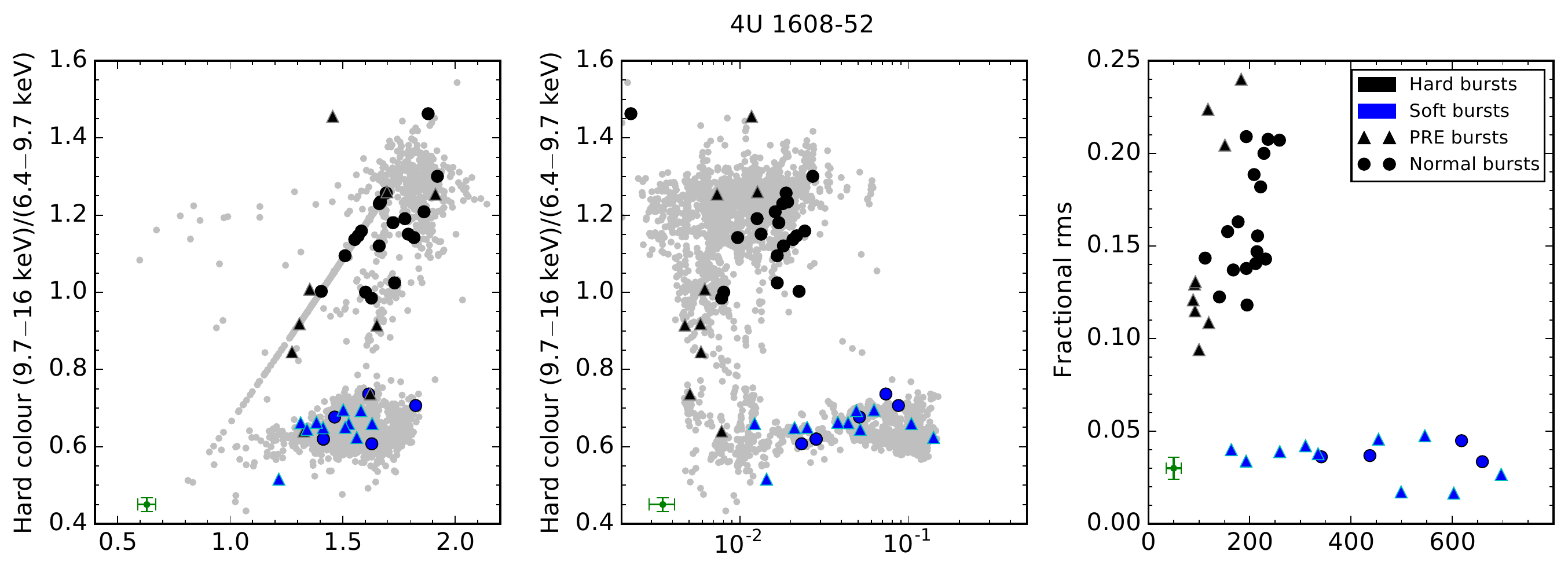}
\includegraphics[width=0.99\textwidth, height=2.5in]{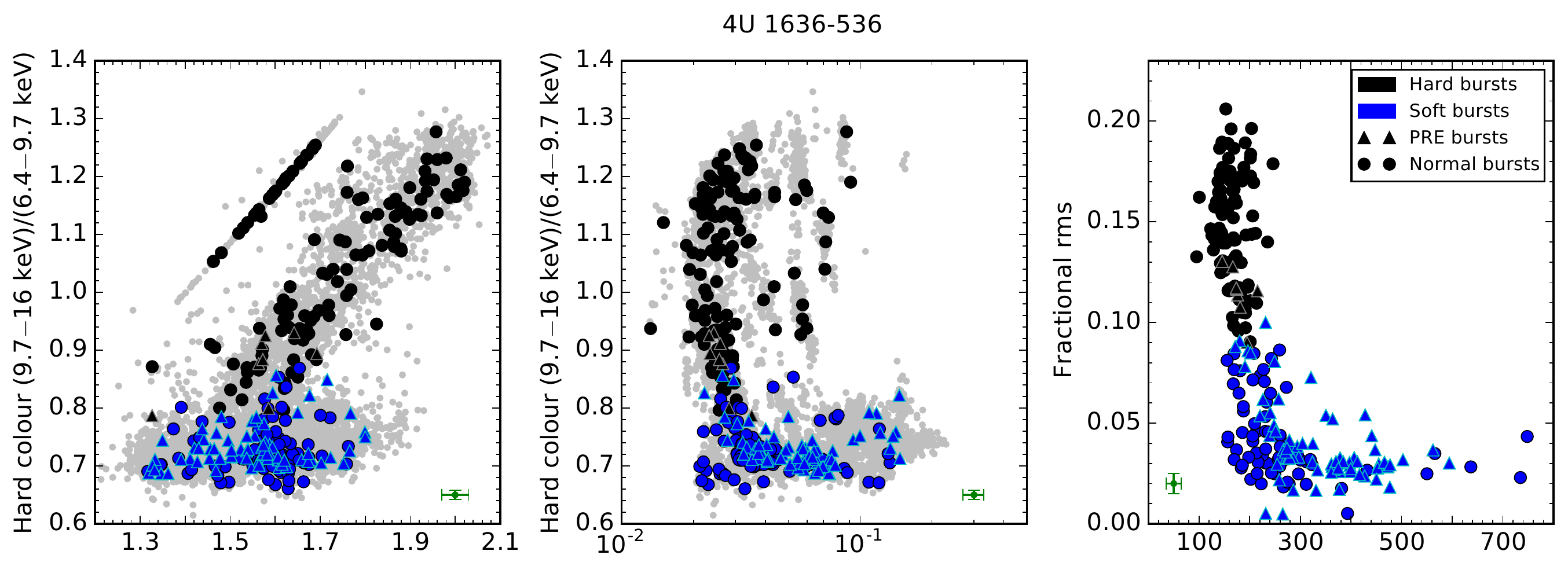}
\includegraphics[width=0.99\textwidth, height=2.5in]{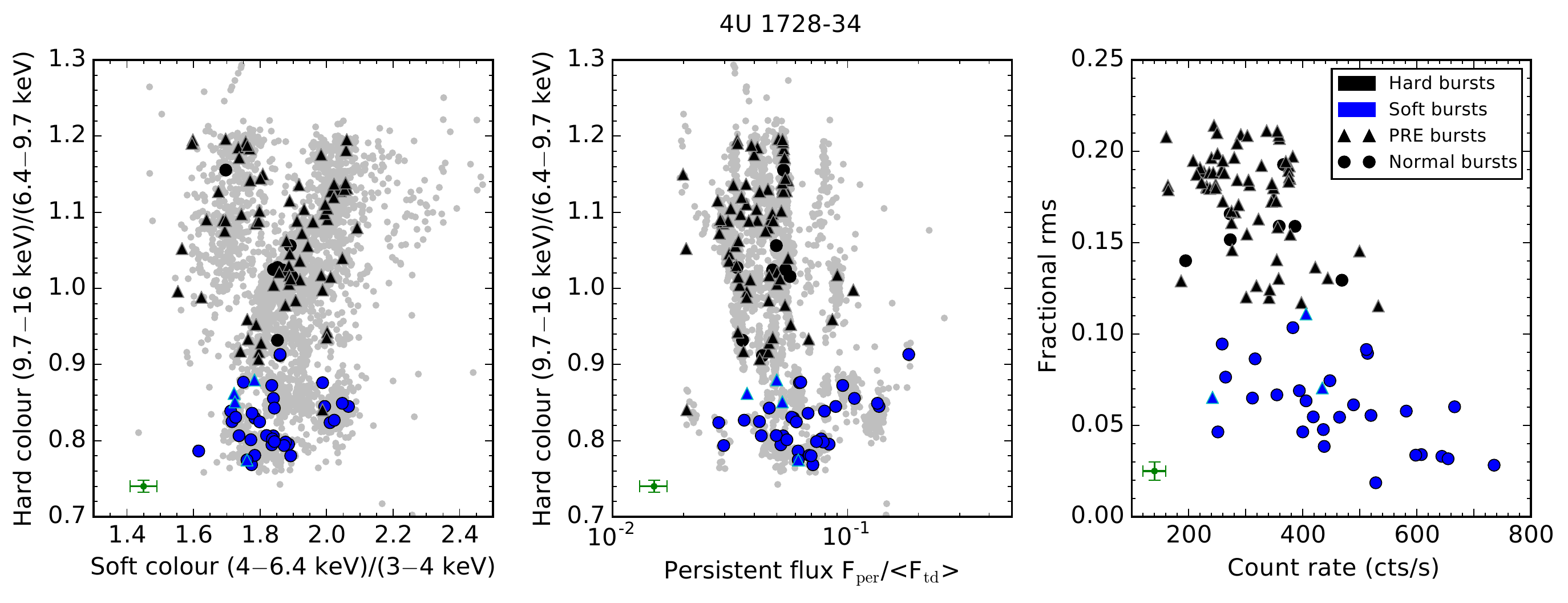}
\end{center}
\caption{State separation for 4U 1636--536, 4U 1608--52, and 4U 1728--34. The left side panels are the colour-colour diagrams from \citet{Kajava14} with the soft and hard state bursts indicated with blue and black markers, and PRE bursts and normal bursts marked with triangles and circles, respectively. The diagonal lines in the left panels are due to our energy band selection and the fact that the persistent emission can sometimes be described with a power law \citep[see][for details]{Kajava14}. In the middle panels, the persistent flux normalised with the average touchdown flux is on the x-axis and the hard colour is on the y-axis (see \citealt{Kajava14}). The right-hand panels show the dependence of the rms on the count rate used to separate the bursts into two states, as explained in Sect.~\ref{sect:states} (see also \citealt{MDFM14}). The green crosses in the bottom of each panel  show the typical error bars. } 
\label{fig:state_separation}
\end{figure*}

\section{Observations}
\label{sec:obs}

\subsection{The data}

We obtained all available public \textit{Rossi X-ray Timing Explorer}/Proportional Counter Array  (\textit{RXTE}/PCA)
data from the High Energy Astrophysics Science Archive Research Center\footnote{http://heasarc.gsfc.nava.gov} (HEASARC) 
for the sources 4U 1636--536, 4U 1608--52, and 4U 1728--34.
We selected these sources because of 
the large number of bursts observed from them by RXTE (about 300 from 4U 1636--536, 120 from 4U 1728--34, and 50 from 4U 1608--52) and because they show both PRE and non-PRE bursts in both the hard and soft states, which allows us to investigate the effect of the spectral state on the X-ray burst decay rates.
We included also 3A 1820--303 (=4U 1820--30) in this work, because it is a well-known ultra-compact binary accreting helium-rich material from its companion \citep{Stella87}, which produces only photospheric expansion bursts in the hard state \citep{GZM13}. 
Thus, this interesting comparison source helps us to understand the effects of spectral state and chemical composition on the cooling rates. In addition, we included GS 1826--24 because of its very regular bursting behaviour and known hydrogen-rich composition of the accreted material \citep{GCK04}.

The X-ray burst spectral data were already used in \citet{Poutanen14} and \citet{Kajava14}, where the data reduction method is described in detail. 
As in those papers, we extracted time-resolved spectra using initial time resolution of 0.25, 0.5, 1.0, or 2.0 s based on the peak count rate (>6000, 6000--3000, 3000--1500, or <1500 count per second, respectively). Then we doubled the time resolution every time the count rate decreased by a factor of $\sqrt{2}$ to maintain the same signal-to-noise ratio.
We modelled the burst emission using a blackbody spectrum ({\sc bbodyrad} model in {\sc xspec}) modified by interstellar absorption ({\sc tbabs}). We fixed the absorption column density $N_\textrm{H}$ to constant values 
$0.16 \times 10^{22}$ cm$^{-2}$ for 3A 1820--303 \citep{Kuulkers03}, 
$0.40 \times 10^{22}$ cm$^{-2}$ for GS 1826--24 \citep{intZand99},
$2.60 \times 10^{22}$ cm$^{-2}$ for 4U 1728--34 \citep{Dai06},
$0.25 \times 10^{22}$ cm$^{-2}$ for 4U 1636--536 \citep{Asai00}, and 
$0.89 \times 10^{22}$ cm$^{-2}$ for 4U 1608--52 \citep{Keek08}.
The best-fitting parameters were then the blackbody temperature $T_{\mathrm{bb}}$ and the normalisation constant $K_{\mathrm{bb}} = (R_{\mathrm{bb}} [\mathrm{km}] / d_{10}) ^2 $, where $d_{10} \equiv d / 10\,\mathrm{kpc}$ is the distance in units of $10$ kpc.
The blackbody flux $F_\textrm{bb}$ was computed using the \textsc{cflux} convolution model over the 0.01--100 keV band, and the parameter errors were calculated as $1\sigma$ confidence levels. We adopted the same criteria described in \citet{Galloway08} to determine if a burst showed signs of PRE, i.e. photospheric radius expansion was considered to occur when the blackbody temperature evolution showed a characteristic double-peaked structure and the blackbody normalisation reached its maximum at the same time as the temperature was at its minimum.

Following \citet{Poutanen14} and \citet{Kajava14}, we extracted a 16-second spectrum prior to the burst and then subtracted it as a background for each burst \citep{Kuulkers02,Galloway08}. While this standard approach is not completely accurate, as the persistent flux might change during the bursts \citep[see e.g.][]{vanParadijs86, Sztajno86, Kuulkers02, Worpel13}, such flux variations are not expected to be as significant in the cooling tails as during the other phases of the burst \citep[see figures 2 and 6 of][]{Worpel13}. Therefore, this operation does not significantly affect the results of our analysis.

\subsection{Accretion states}
\label{sect:states} 

The accretion states can be defined based on the X-ray hardness (see e.g. \citealt{Hasinger1989}) and on the total fractional rms of the variability \citep{Munoz-Darias2011}. 

\citet{MDFM14} defined three main spectral states in accreting NS X-ray binaries, based on the total fractional rms: a hard state at rms > 20\%, an intermediate state at rms between 5 and 20\% rms, and a soft state at rms < 5\%. However, these authors note that different sources might span different rms ranges and the state classification given above has to be adjusted. On the other hand, X-ray colours have been historically used to determine the spectral states of accreting NS, but while in certain sources the distinction between hard and soft states (i.e. roughly corresponding to the so-called island state and the banana state of \citealt{Hasinger1989}) is clear, in others it remains somewhat arbitrary \citep{Kajava14}. 
 
In order to quantify a possible dependence of the burst flux decay on the spectral state, we classified the bursts of our sample into two different groups based on the spectral state during which they were observed. The least ambiguous way to determine the spectral state is using both the rms and colour information (see Fig.~\ref{fig:state_separation}). 
We obtained X-ray colours and persistent fluxes following \citet{Kajava14} in the 3--4 to 4--6.4~keV energy band (soft colour) and 6.4--9.7 to 9.7--16~keV energy band (hard colour). We measured the rms following \citet{MDFM14} in the 2--16~keV energy band and in the 0.1--64~Hz frequency band. Both the colours and rms were measured from the persistent emission before the onset of each burst. 

Comparing the state classification defined by \cite{MDFM14} and the X-ray colour analysis reported by \citet{Kajava14} for 4U 1636--536, 4U 1608--52, and 4U 1728--34, we found that for a few bursts the two methods resulted in ambiguous classifications (particularly for 4U 1636--536), and in those cases we adjusted the limits so that both colour and rms diagrams had the least ambiguities.
We also find that (see Fig.~\ref{fig:state_separation})

\begin{itemize}
\item Most of the bursts found at rms lower than 9\% are produced in the soft state. We refer to these bursts as \text{soft bursts}.
\item The majority of the bursts observed above 9\% rms are generally found in either the intermediate or hard state. We generically refer to these bursts as \text{hard bursts}. 
\end{itemize}

\begin{figure*}[!ht]
\begin{center}
\includegraphics[width=0.99\textwidth]{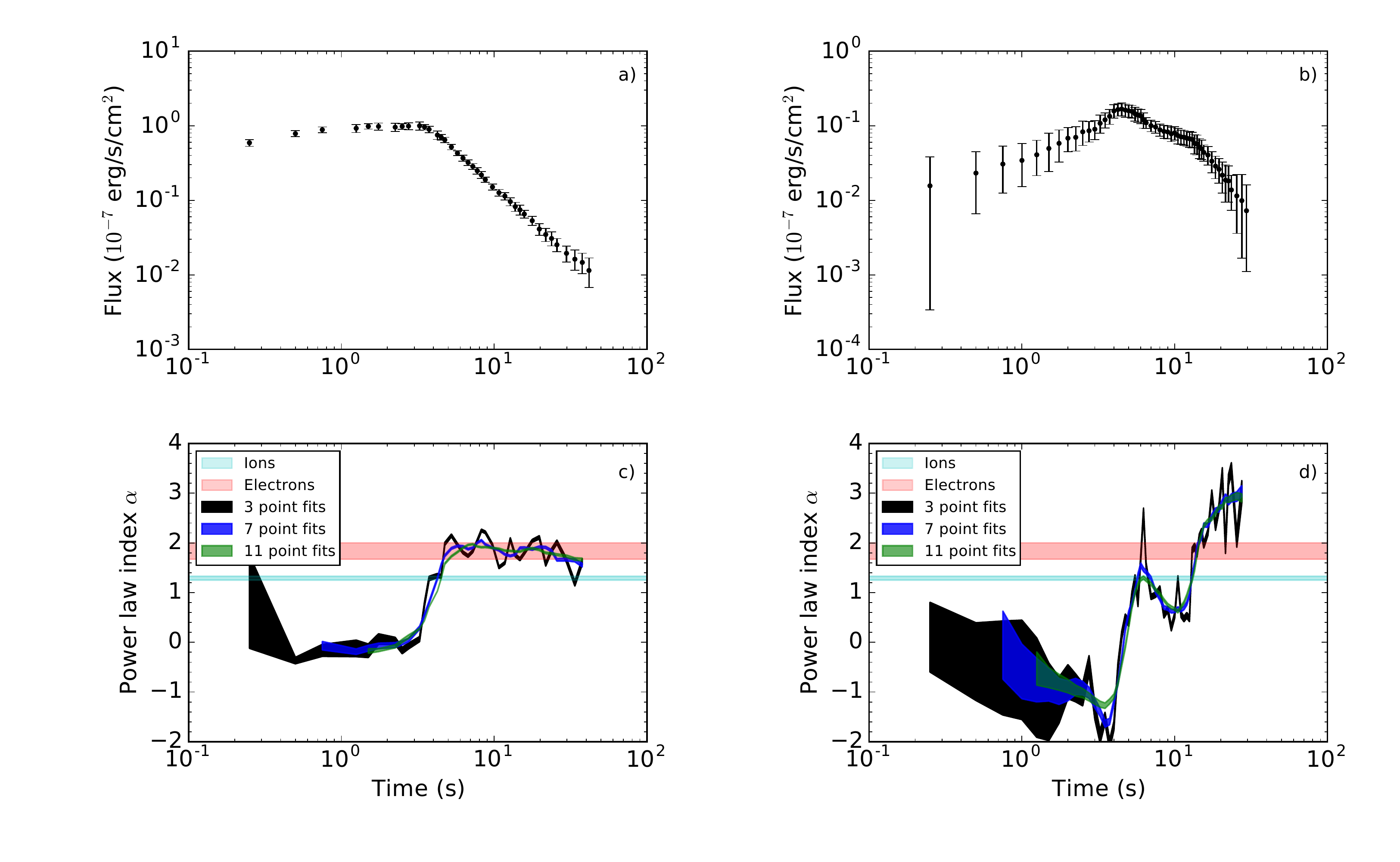}
\end{center}
\caption{Two examples of application of the dynamic power-law index method to the bursts.  The upper panels show the temporal evolution of the flux, while the lower panels give power-law indices. 
The left panels represent a PRE burst from 4U 1728--34 (MJD 51238.79155),  which shows 
a nearly power-law dependence of flux on time; the power-law index $\alpha$ varies smoothly (panel~c). 
The right panels represent a non-PRE burst from 4U 1636--536 (MJD 52290.14847), whose temporal evolution does not follow a power law with the corresponding power-law index shown in panel~d. 
In the lower panels the black, blue, and green areas represent the 1$\sigma $ confidence levels of the power-law indices of the individual 3, 7, and 11 point fits, respectively.
The power-law indices are shown for the entire burst, but we study only the cooling tails.
Also in the lower panels the red area shows the typical $\alpha$-regime if the cooling is dominated by electrons, and the cyan area corresponds to ion-dominated cooling (see text). 
}
\label{fig:powfit}
\end{figure*}

\section{Analysis}\label{sec:anal}

\subsection{Theory}\label{sec:theory}

It has been customary to model the cooling phase of an X-ray burst with a simple exponential decay (see e.g. \citealt{Grindlay76b,Galloway08}). 
While some bursts can be described with exponential decay, it does not fit well all the bursts nor is it based on correct physical considerations. 
By considering one-dimensional radiative heat transfer and the Stefan-Boltzmann law, \citet{intZand14} showed that the burst luminosity should follow a power-law-like decay 
\begin{equation}
L \propto t^{-\alpha},
\end{equation}
where the power-law index $\alpha$ equals 4:3 if the dominant contribution to the heat capacity comes from ions, $\alpha =2$ if it is dominated by degenerate electrons, or $\alpha > 2$ if the heat capacity is dominated by the radiation field. These cooling regimes are illustrated in Fig.~\ref{fig:powfit}c and Fig.~\ref{fig:powfit}d. 
The bursting NS envelope is a mixture of all these components, so the decay index should lie somewhere in between these values. \citet{intZand14} have also mentioned that a similar consideration of simple conductive heat transfer produces an exponential decay, but they have found that the power law was always a better representation of the cooling with majority of the bursts having the $\alpha$ between 4:3 and 2. On the other hand, they had selected only a small subset of `clean' and ordinary bursts to get a clear view of the cooling process without the contribution of prolonged nuclear burning, varying accretion rate, or accretion disc.
While many bursts indeed follow the power-law decay very well (Fig.~\ref{fig:powfit}a), there are also bursts that do not, nor can they be fitted with an exponential decay (Fig.~\ref{fig:powfit}b). Models of \citet{Cumming04}, on the other hand, predict that the cooling rate changes during the burst, i.e. the power-law index is varying, which could be expected if the governing component changes with the temperature and flux.

\subsection{Temporal evolution of the flux}

To study the possible changing cooling rate of X-ray bursts, we present a dynamic power-law index method, where within a moving time window we fit a power-law function of time to the flux as follows:   
\begin{equation}
\label{pow}
F(t) \approx F_{0} \ (t/t_0) ^{-\alpha}, 
\end{equation}
where $t$ is the time in seconds, $F_{0}$ is a free parameter (whose value is usually close to the flux of the first data point of the window $t_0$), and $\alpha $ is the power-law decay index\footnote{Instead of assuming a separate $\alpha$ value for each bin, we could also try to capture its evolution by assuming some a priori functional form for the function $\alpha(t)$ and try to constrain that using the information from the whole burst tail. 
However, this kind of formulation does not allow us to examine any unexpected or irregular cooling behaviour that we observe here.} 
\begin{equation}
\label{alpha}
\alpha = - \frac{d\log F}{d\log t}.  
\end{equation}
Low values of $\alpha$ correspond to a slow local change in the flux, whereas high values of $\alpha$ indicate a fast change.

We adopted a time window of seven bins and fitted a power law to the measured burst blackbody flux values in this narrow time range, then moved the window one step forward and fitted again and so on until the end of the burst, which we identify when typically a flux $\sim$5\% of the peak flux is reached. We fitted the whole burst, but consider only the cooling part, where the $\alpha$ is positive. The first fits with positive $\alpha$ are still affected by the peak of the burst, which means that at first the $\alpha$-values rise rapidly before settling to the actual decay values.  

The effect of the selected window size on the results was also investigated and was found to have no impact on the underlying $\alpha$-evolution, within a reasonable time range, as shown in Fig.~\ref{fig:powfit}c and~\ref{fig:powfit}d, which provide $\alpha$ as a function of time for windows of 3, 7, and 11 point fits.
However, a narrower window produced noisier results, while larger number of points tends to smooth the $\alpha$-flux evolution too much. 
As a good balance between these two effects we found that a 7 point fit produces the best result for our data reduction and bin summation methods (see Sect.~\ref{sec:obs} and also \citealt{Poutanen14} and \citealt{Kajava14}).

\subsection{$\alpha$-flux evolution}

In Figs.~\ref{fig:1820}--\ref{fig:1608} we show the values of the power-law index $\alpha$ obtained from the moving time-window fits as a function of the burst flux. 
These figures show how $\alpha$, i.e. the cooling rate, changes as the burst flux decays. 
In these figures the peak of the burst is on the right-hand side and the flux decreases from right to left. 
The bursts are grouped based on the accretion state and whether or not they are PRE bursts.

In Figs.~\ref{fig:1820}--\ref{fig:1608} we also show the burst cooling models adapted from \citet{intZand14}. 
These models include their one-zone model with an ignition depth of $10^{8}$ g cm$^{-2}$ and three multi-zone models with ignition depths of $10^{8}$, $10^{9}$, and $10^{10}$ g cm$^{-2}$, which are based on the work of \citet{Cumming04}. 
The fluxes provided by these models were arbitrarily scaled to coincide with the fluxes measured for each source.
In practice this scaling connects the emergent and observed flux from the source,  
\begin{equation}
\label{em_flux}
F =  \frac{F_{\rm NS}}{(1+z)^2} \frac{R^2}{D^2} ,
\end{equation}
where $F$ is the observed flux, 
$F_{\rm NS}$ is the flux at the NS surface,  $R$ is the NS radius, $1+z = (1-2GM/R c^2)^{-1/2}$ is the redshift, $G$ is the gravitational constant, $M$ is the NS mass, $c$ is the speed of light, and $D$ is the distance to the source. 
The scaling applied to the models is consistent within a factor of 2--3 with what one would expect from Eq.~(\ref{em_flux}) for a canonical NS ($M = 1.4 M_{\odot}$, $R$ = 10 km) with distances 5.3 $\pm$ 0.8 kpc for 4U 1728--34 \citep{Jonker04}, 6.0 $\pm$ 0.5 kpc for 4U 1636--536 \citep{Galloway06}, 7.6 $\pm$ 0.4 kpc for 3A 1820--303 \citep{Heasley00}, 3.4 $\pm$ 0.3 kpc for 4U 1608--52 \citep{Poutanen14}, and 5.7 $\pm$ 0.2 kpc for GS 1826--24 \citep{Chevenez16}.
Here the distance to the source is by far the largest origin of error in the scaling. 

In the following Sections \ref{sec:1820}--\ref{sec:1608} we describe the results of each source in detail. In this work we focus mostly the $\alpha$-flux evolution of the bursts, even though both the flux and $\alpha$ are also functions of time. One should keep in mind that even a small change in $\alpha$-flux evolution can mean a large change in, for example the burst duration.

\subsubsection{3A 1820--303}
\label{sec:1820}

\begin{figure}
\begin{center}
\includegraphics[width=0.55\textwidth]{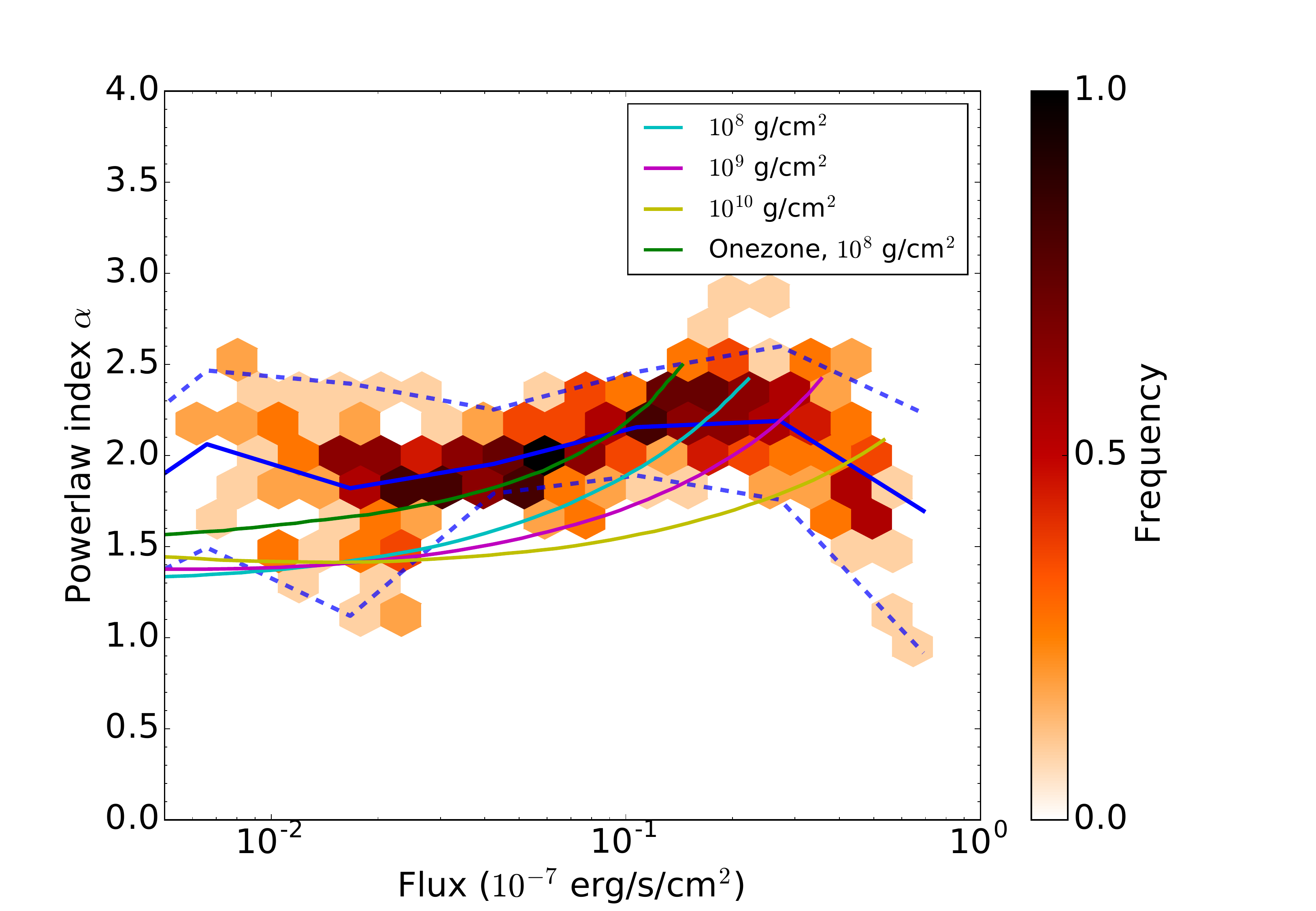}
\end{center}
\caption{Dependence of the power-law index $\alpha$ on the burst flux for all bursts from 3A 1820--303 (all of these bursts are hard-state PRE bursts). 
The data are shown in a two-dimensional histogram, where the colour of a bin shows the frequency of occurrences (normalised to the peak value) according to the colour bar on the right.  
The solid blue line is the average of the data and the blue dashed lines show the 1$\sigma$ limits. The solid green line shows the one-zone  model with a column depth of $10^{8}$ g cm$^{-2}$, and the cyan, magenta, and yellow lines show the three multi-zone models with ignition depths of $10^{8}$, $10^{9}$, and $10^{10}$ g cm$^{-2}$, respectively, based on the work of \citet{Cumming04}. 
} 
\label{fig:1820}
\end{figure}

3A 1820--303 (= 4U 1820--30) (Fig.~\ref{fig:1820}) is an ultra-compact binary with an orbital period of just 11.4 min \citep{Stella87}. The companion star is a low-mass helium white dwarf, which means that the accreting material and thus the burning material on the surface of the NS, is mainly helium \citep{Rappaport87}. In 3A 1820--303 all the bursts occur in the hard state and show signs of photospheric radius expansion. 

From Fig.~\ref{fig:1820} we can see that all the bursts from this source follow the same behaviour. After the peak of the burst, $\alpha$ rises quickly to about 2.2, after which it starts to slowly decrease with the flux being around 2 most of the time. 
This means that the cooling is consistent with heat transfer dominated by electrons. 
At low fluxes, $\alpha$ seems to rise again, while at the same time the scatter increases, which is likely due to the burst flux approaching the persistent emission level. 
There is a qualitative similarity between the models and data, but the models predict lower values of $\alpha$, i.e. slower cooling, than we see in the data, especially at lower fluxes. A similar discrepancy between the cooling rate of models and data is also seen in a superburst from 4U 1636--536 (\citealt{Keek15}; but see also \citealt{Koljonen16}).

\subsubsection{GS 1826--24}
\label{sec:1826}

\begin{figure}
\begin{center}
\includegraphics[width=0.55\textwidth]{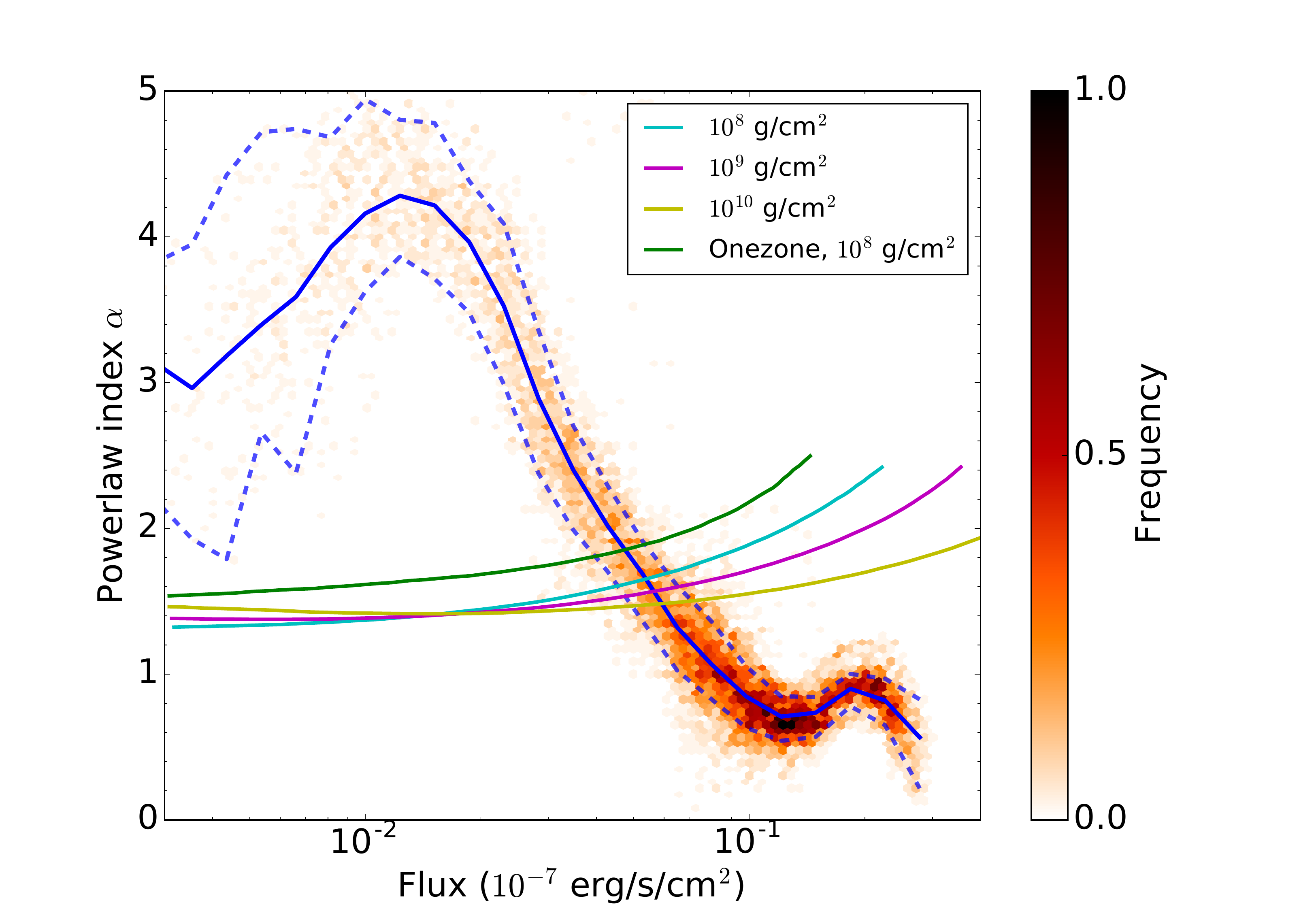}
\end{center}
\caption{Same as Fig.~\ref{fig:1820}, but for GS 1826--24. } 
\label{fig:1826}
\end{figure}

GS 1826--24 (also known as Ginga 1826--238) is a very extensively studied burster because of its bright, frequent, and regular bursts. \citet{Ubertini99} dubbed it the `clocked burster' because of the regular bursting behaviour, while \citet{Bildsten00} name it the `textbook burster' because of the good agreement with theory. 

\citet{Bildsten00} suggested that the bursts involve mixed H/He burning, where an initial helium flash prompts prolonged hydrogen burning via the rp-process \citep{Wallace81}.
This naturally explains the observed long tails ($\approx 100$ s) of the bursts \citep{HCG07}. 
\citet{GCK04} arrived at the same conclusion by showing that the regular bursting behaviour is well understood as being due to He ignition in a H-rich environment. The inferred accretion rate $\dot{M} \approx 10^{-9} M_{\odot}$ yr$^{-1}$ implies that hydrogen burns also stably between the bursts by the beta-limited hot CNO-cycle \citep{GCK04}.

The regular and very well understood behaviour makes GS 1826--24 an ideal calibration source for hydrogen rich bursts. 
In Fig.~\ref{fig:1826} we show the results of the dynamic power-law method for this source. 
From this figure we can see that all the bursts indeed follow closely the same behaviour, where the $\alpha$ first rises quickly to about 1, after which it first slightly decreases to around $\alpha \approx 0.8$ and then rises dramatically to all the way to $\alpha \approx 4$. 
The rise in $\alpha$ is almost linear, which means that the flux decays almost exponentially, since in the $\alpha$ versus flux space an exponential function is a straight inclined line. 
At low burst fluxes the $\alpha$ decreases again, but the scatter is substantial at the end of the cooling tail.

\subsubsection{4U 1728--34}
\label{sec:1728}

\begin{figure*}
\begin{center}
\includegraphics[width=0.75\textwidth]{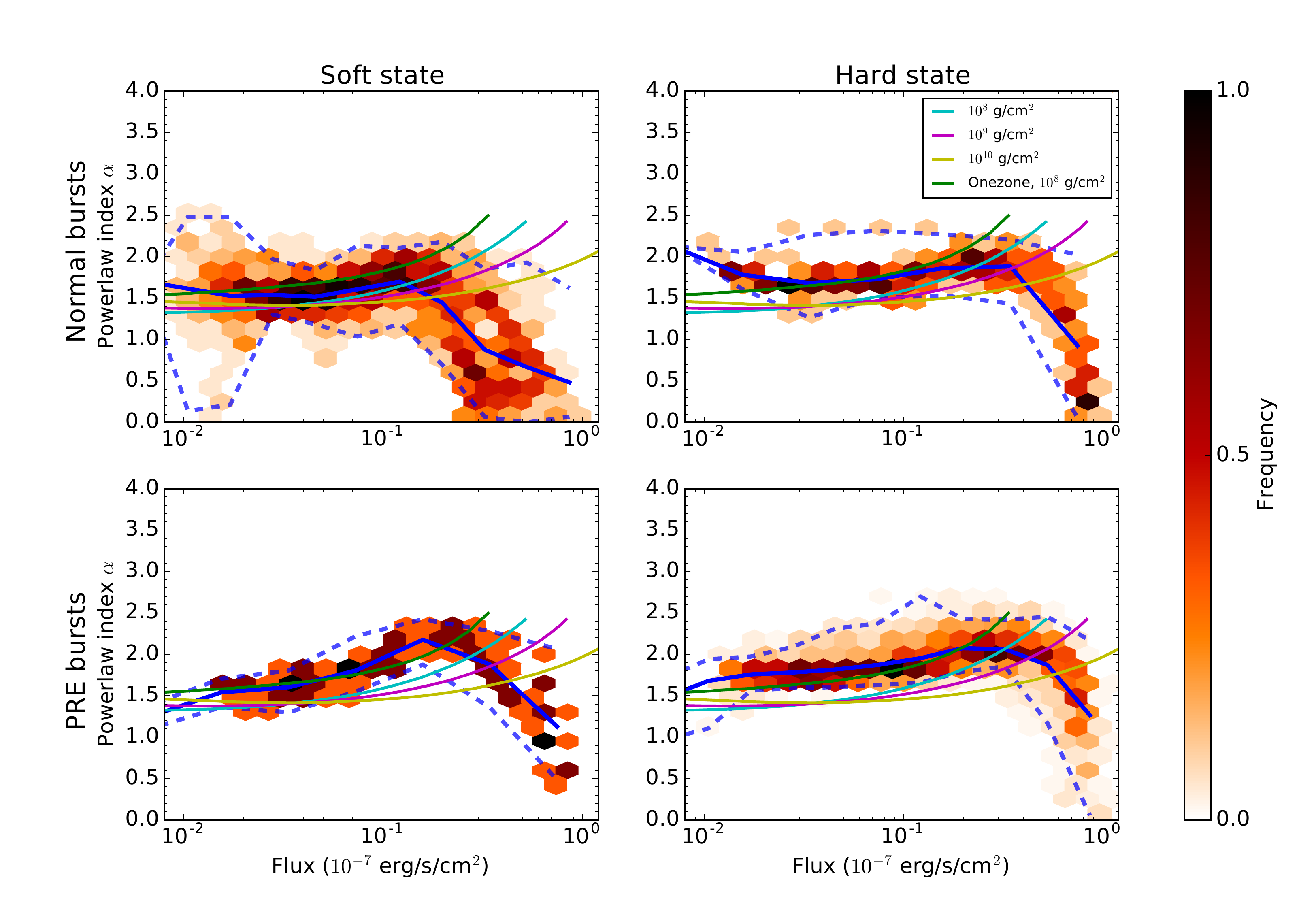}
\end{center}
\caption{Same as Fig.~\ref{fig:1820}, but for 4U 1728--34.
The data are shown for PRE bursts (lower panels) and normal bursts (upper panels) in the hard (right panels) and soft state (left panels). } 
\label{fig:1728}
\end{figure*}

4U 1728--34 (Fig.~\ref{fig:1728}) is also a H-poor ultra-compact binary candidate \citep{Galloway10}, and the burst properties indeed closely resemble those of 3A 1820--303 \citep[see e.g.][]{Galloway08}. 
However, unlike 3A 1820--303, 4U 1728--34 does have bursts both in the hard and soft state and not all of the bursts show a PRE stage. 
In 4U 1728--34 both the normal and PRE bursts cool down approximately in the same way as in 3A 1820--303, regardless of the accretion state. 
The non-PRE bursts in the soft state just have slightly lower peak fluxes and more burst-to-burst variations because they are fainter.
The bursts from this source show a clear similarity to the models with the low-flux decay indices closer to the predicted value of $\alpha \sim 1.4$ than in 3A 1820--303.

\subsubsection{4U 1636--536}
\label{sec:1636}

\begin{figure*}
\begin{center}
\includegraphics[width=0.75\textwidth]{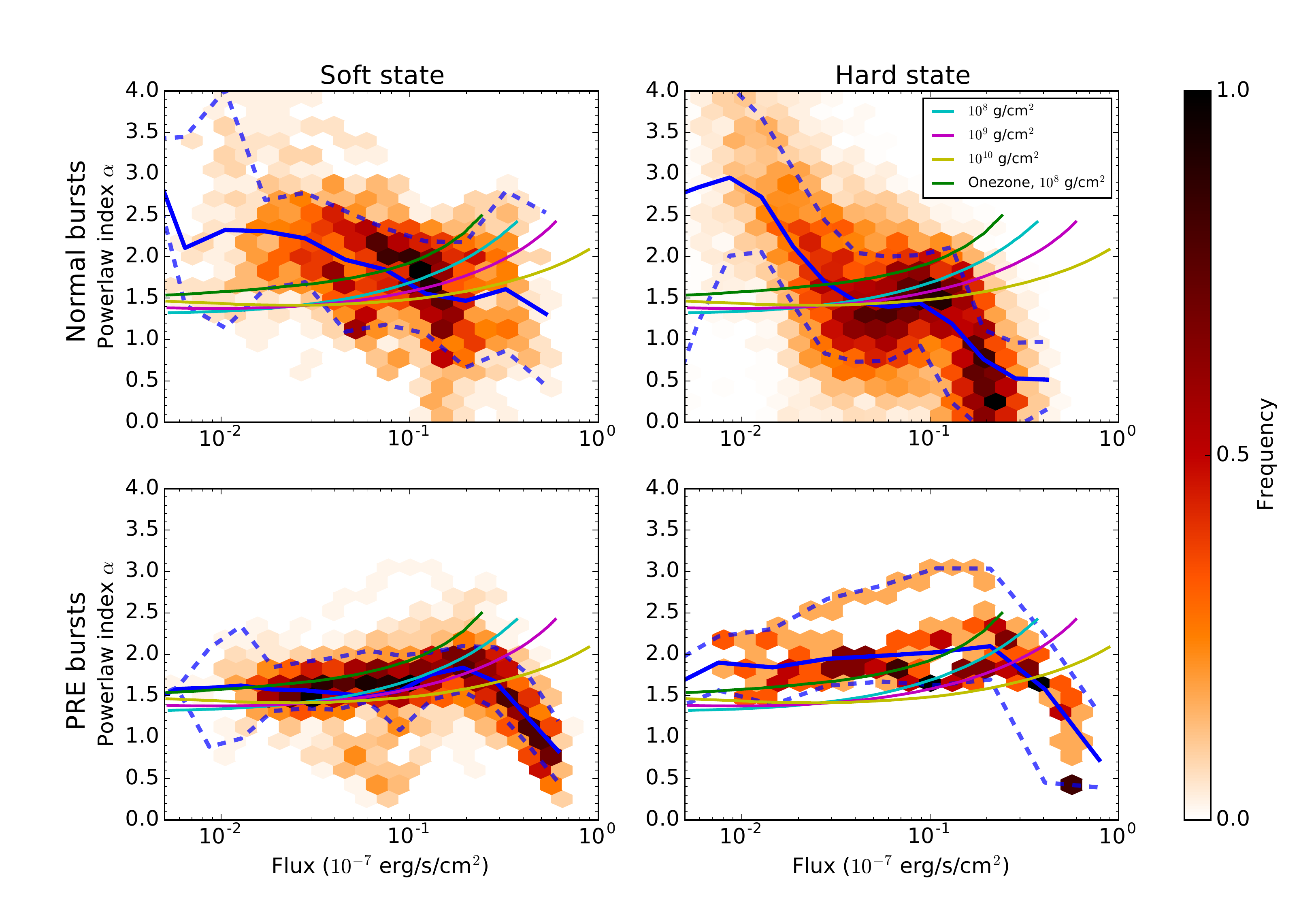}
\end{center}
\caption{Same as Fig.~\ref{fig:1728}, but for 4U 1636--536. } 
\label{fig:1636}
\end{figure*}

4U 1636--536 (Fig.~\ref{fig:1636}) is believed to accrete both H and He because its 18 mag optical counterpart V801 Arae has 3.8~h orbital period and clear hydrogen emission lines \citep{vanParadijs90, AvdHdJ98, Giles02}. 
In this source all the PRE bursts except six happen in the soft state and even these six happen during the transition from one state to the other and they resemble the soft state more than the actual hard state bursts.  
All the PRE bursts in this source (lower panels of Fig.~\ref{fig:1636}) closely resemble the bursts from 3A 1820--303 and 4U 1728--34. 
They also qualitatively follow the models, but there are some bursts with clearly higher and lower $\alpha$-values.

The hard state normal bursts in 4U 1636--536 (upper right panel of Fig.~\ref{fig:1636}) clearly differ from the PRE bursts. 
The index $\alpha$ first rises quickly when the burst starts to cool and then it starts to decrease slowly as in the PRE bursts.
At lower fluxes, however, $\alpha$ starts to rise again reaching values as high as $\alpha \approx 4$, i.e. the bursts cool down much more rapidly than predicted by the cooling models.
This behaviour is similar to GS 1826--24, but whereas the $\alpha$ in GS 1826--24 stays below 1 in the beginning of the cooling, in 4U 1636--536 it rises at first close to 2 before stalling and rising again. Overall, the shape of the $\alpha$-flux curve of the hard state normal bursts is close to that of GS 1826--24, but the actual values differ.

The normal bursts that occur in the soft state do not follow the behaviour of either of the two calibration sources. 
Overall, these bursts lack the decrease of $\alpha$ altogether, seen in the other bursts and sources.
The burst flux decay seems to intensify almost monotonically as the atmosphere cools down. 
Also, the effects in the tail are not due to the background emission subtraction as we carefully tested different data reduction schemes but still obtained the same behaviour.

\subsubsection{4U 1608--52}
\label{sec:1608}

\begin{figure*}
\begin{center}
\includegraphics[width=0.75\textwidth]{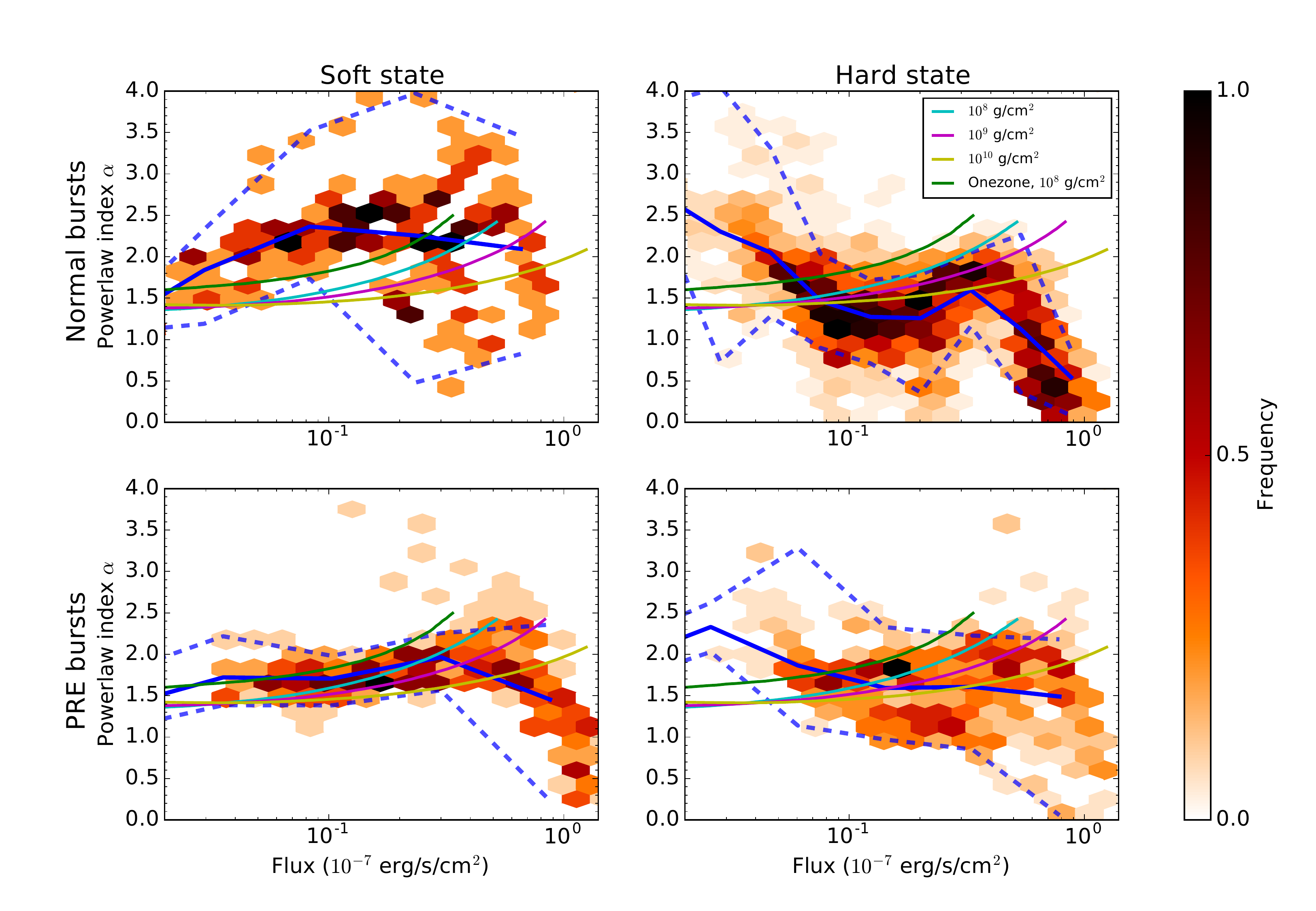}
\end{center}
\caption{Same as Fig.~\ref{fig:1728}, but for 4U 1608--52. }  
\label{fig:1608}
\end{figure*}

Similar to 4U 1636--536, 4U 1608--52 (Fig.~\ref{fig:1608}) is also believed to accrete both H and He, since its optical counterpart QX Nor is likely a late F- or an early G-type main sequence star with an orbital period of 0.537~d \citep{Wachter02}.
The normal bursts of 4U 1608--52 in the hard state are similar to those of 4U 1636--536, having the peculiar rise of $\alpha$ at low fluxes, but in this source the $\alpha$ rises only to around 2.5 compared to the $\alpha$ $\approx$ 4 in 4U 1636--536.
However, in the soft state normal bursts there does not seem to be any rise in the $\alpha$-values at low fluxes, but instead they nearly monotonically decrease during the decay.

The PRE bursts that occur in the soft state clearly follow the same behaviour as the PRE bursts in 4U 1636--536 and all the bursts in 4U 1728--34 and 3A 1820--303. 
Unlike in 4U 1636--536, 4U 1608--52 has several PRE bursts in the hard state and they are different from the soft state bursts. 
There is a clear rise of $\alpha$ at low fluxes resembling the normal bursts in the hard state and also the decay indices show larger scatter than in the soft state PRE bursts.

\section{Discussion}\label{sec:disc}

The type-I X-ray burst decays have been traditionally described with a single exponential function and recently also with a single power law \citep{intZand14}. 
Our results show that, while the flux decay during some of the X-ray burst cooling tails can indeed be approximated with a power law, clearly not all the bursts can be fitted with just one constant power-law decay index $\alpha$ because the cooling rate changes in time (see Figs.~\ref{fig:1820} --~\ref{fig:1608}). This is also predicted by the cooling models of \citet{Cumming04}.

Burst cooling is more complex than traditionally thought. 
As one can see from Figs.~\ref{fig:1820} --~\ref{fig:1608}, there are differences in the cooling between different sources and burst types; the most notable difference is the rise in the cooling rate at lower fluxes.
As the simplest case, we can consider 3A 1820--303, which is known to be an ultra-compact binary accreting mainly helium \citep{Stella87}.
All of the bursts from this source are thought to be produced by unstable burning of helium; we refer to these bursts as helium bursts. 
These bursts are found to be qualitatively similar to the simple cooling models of \citet{Cumming04} and \cite{intZand14}.  
Similarly, 4U 1728--34 is also believed to be an ultra-compact binary \citep{Galloway10}, and comparing Fig.~\ref{fig:1820} and Fig.~\ref{fig:1728} certainly supports this notion because the $\alpha$-flux evolution is very similar.
All the bursts from 4U 1728--34, regardless of the state or burst type, seem to follow the cooling models, even better than the bursts from 3A 1820--303. 
Comparing these two He-rich sources and the cooling models, we can argue that the helium bursts can be qualitatively described by the cooling models of \citet{Cumming04}. 
One reason for this is that these models, initially constructed for superbursts, assume instantaneous injection of energy, which is a reasonable assumption for the helium bursts that rapidly exhaust the available fuel via the fast triple-alpha process.

Hydrogen, on the other hand, burns much slower via $\beta$-decay limited hot CNO-cycle \citep{Fowler65} and rp-process \citep{Wallace81}, so the energy is released slower. 
How all this then affects the evolution of $\alpha$ is most evident in GS 1826--24, which is, in contrast, known to be a H-rich source with prominent rp-burning tails \citep{GCK04,HCG07}.
The cooling of bursts in GS 1826--24 is very different from the He-rich sources. 
The cooling tail of these bursts can be divided into three phases. 
First the $\alpha$ stays roughly constant at about 1, while the hydrogen is still burning via the rp-process \citep{HCG07}. 
After that the cooling rate increases almost linearly to $\alpha \approx 4,$ meaning that the flux decays exponentially in contrast to the power-law-like cooling of the He-rich sources. 
The final third phase of the decay at the very end of the burst, where the $\alpha$ seems to decrease, may be the onset of the one-hour long tails observed by \citet{intZKC09}, which can be explained by delayed cooling of deeper layers that were heated up through inward conduction.

The flux decay behaviour of 4U 1636--536 and 4U 1608--52 is more complex than that of the previous sources, as there are differences between the accretion states.
All the PRE bursts of 4U 1636--536 and the soft state PRE bursts of 4U 1608--52 are similar to the bursts of the two helium sources, and they also follow the simple cooling models. 
These bursts support the argument of \citet{Ebisuzaki88} and \citet{Zhang11}  that PRE bursts are H poor.
On the other hand, the hard state PRE bursts of 4U 1608--52 differ. 
They are similar to the hard state normal bursts because in both of these  the power-law index $\alpha$ rises at lower burst fluxes as  normal bursts of 4U 1636--536 do in the hard
state. 
The increase in $\alpha$ is similar to GS 1826--24, which is known to be in the "Case III" burning regime of \citet{NH03} producing mixed H/He bursts with stable H burning in between, but the $\alpha$-evolution is still quantitatively different.
In the bursts from GS 1826--24 the $\alpha$ is at first less than one, i.e. there is hardly any cooling at all, possibly because of the rp-burning.
In both 4U 1636--536 and 4U 1608--52 the $\alpha$ is higher, around 1.5, indicating that the cooling is faster. 
The rise is the most prominent in GS 1826--24 and most subtle in 4U 1608--52. 
This increase in the cooling rate at lower burst fluxes is seen only in these three supposedly H-rich sources.
The differences could arise because of the different amount of available hydrogen, variations in metallicity, or in the area on which new fuel accretes \citep{Bildsten00}.

The situation of the soft state normal bursts of 4U 1636--536 and 4U 1608--52 is more complex.
All soft state bursts have been suggested as H poor \citep[e.g.][]{Zhang11}, and 4U 1608--52 supports this notion, to some extent, since all the soft state bursts are similar to 3A 1820--303. 
In 4U 1636--536 the $\alpha-F$ evolution of the soft state normal bursts seems to be somewhere in between the H-poor PRE and the H-rich hard state normal bursts. 
This could be partly because in 4U 1636--536 the accretion states are not as well defined as in 4U 1608--52 (see Fig.~\ref{fig:state_separation}) and we have neglected the intermediate state that can, in fact, be more similar to the soft state than to the hard state.

Our results show that while the helium bursts can be qualitatively described with the simple models of \citet{Cumming04}, the mixed bursts display more complex behaviour. 
On the other hand, \citet{HCG07} showed that the burst light curves of GS 1826--24 can be well reproduced with models of \citet{Woosley04}, which include a large nuclear reaction network to follow the rp-process. But there are also some differences between the models, depending on the nuclear physics and the physical treatment. 
This is evident in Figs.~\ref{fig:1820} --~\ref{fig:1608} as well, where the one-zone and multi-zone models produce different results \citep[see also][]{CAH16}

The most comprehensive, and also the most recent, modelling of bursting behaviour has been carried out by \citet{Lampe16}, who used the {\sc kepler} hydrodynamic code to simulate bursts with varying accretion rates across a range of chemical compositions; these simulations also included the He-rich and solar-like H/He ratios, which are similar to the sources in our work.
A detailed comparison with these models and the data is therefore in order.

\citet{Lampe16} found that the accretion rate is one of the governing factors of bursting properties and burst morphology. 
Our results support this conclusion. 
The accretion state seems to affects the burst cooling, as we observe mixed H/He bursts with the rise in the power-law index only in the hard state (low persistent flux) and pure He bursts in the soft state (higher persistent flux), which is opposite from what \citet{NH03} have predicted.
This may be due to the differences in the accretion geometry between these two states.
In the hard state the accretion flow onto the surface is more likely spherical and consists of hot electrons, protons, and heavier ions that deposit most of their energy at the upper atmosphere layers \citep{BSW92}.
In the soft state, instead, the flow settles onto the neutron star surface through the optically thick boundary/spreading layer, where the flow spreads from the NS equator to higher latitudes depending on the accretion rate \citep{IS99,IS10}. 
Furthermore, the settling of matter onto the neutron star is also expected to cause significant heat dissipation in the burning depths, and thus the formation of a spreading layer in the soft state may cause changes in burning regimes.

In the end, we conclude that studying the cooling of the X-ray bursts with the aforementioned dynamic power-law method can be a very powerful tool in probing the interiors of NSs.
Firstly, we get indirect information about the fuel composition because H-rich and H-poor bursts produce characteristically different cooling shapes. 
Secondly, we can study the burst ignition and burning mechanism because the slope of the flux decay seems to be sensitive to any additional heating, such as the prolonged rp-burning in the NS ocean.
Lastly, it is interesting to see how the actual physical environment of the NS affects and modifies the cooling.
Here the differences in the accretion rate might also play a role because of the clear differences between the bursts occurring in the hard and soft states.
As a next step it is also possible to do a similar analysis to simulated bursts such as those produced by the \textsc{kepler} code \citep[see][]{Lampe16}.
Hence, when the input is known, we can calibrate our method further and study, for example the effects of composition and accretion in more detail.
This is a matter of our follow-up work.

\begin{acknowledgements}
We thank Jean in 't Zand and Duncan Galloway for interesting discussions.
JK, JJEK, and JP acknowledge support from the Academy of Finland grant 268740. 
JJEK also acknowledges support from the ESA research fellowship programme and the Academy of Finland grant 295114. 
JN acknowledges support from the University of Turku Graduate School in Physical and Chemical Sciences. 
JP was partly supported by the Foundations' Professor Pool, the Finnish Cultural Foundation, and the National Science Foundation grant PHY-1125915.
\end{acknowledgements}

\bibliographystyle{aa}
\bibliography{viitteet}

\end{document}